\documentclass[prl,twocolumn,showpacs,amsmath,amssymb,aps]{revtex4}

\usepackage{graphicx}

\begin{document}

\title
{Liquid-like behavior of supercritical fluids}

\author{F.~Gorelli$^{1,2}$, M.~Santoro$^{1,2}$, T.~Scopigno$^1$}
\email{tullio.scopigno@phys.uniroma1.it}
\author{M.~Krisch$^3$}
\author{G.~Ruocco$^{4,1}$}

\affiliation{
$^1$Research center SOFT-INFM-CNR,  Universit\`a di
Roma ``La Sapienza,'' I-00185, Roma, Italy \\
$^2$ LENS, Via N. Carrara 1, I-50019 Sesto Fiorentino, Firenze,
Italy \\
$^3$ European Synchrotron Radiation Facility, BP 220, 38043 Grenoble, France\\
$^4$ Dipartimento di Fisica, Universit\'a di Roma "La Sapienza",
00185 Roma, Italy }

\begin{abstract}
The high frequency dynamics of fluid oxygen have been investigated
by Inelastic X-ray Scattering. In spite of the markedly
supercritical conditions ($T\approx 2 T_c$, $P>10^2 P_c$), the
sound velocity exceeds the hydrodynamic value of about $20 \%$, a
feature which is the fingerprint of liquid-like dynamics. The
comparison of the present results with literature data obtained in
several fluids allow us to identify the extrapolation of the
liquid vapor-coexistence line in the ($P/P_c$, $T/T_c$) plane as
the relevant edge between liquid- and gas-like dynamics. More
interestingly, this extrapolation is very close to the non
metal-metal transition in hot dense fluids, at pressure and
temperature values as obtained by shock wave experiments. This
result points to the existence of a connection between structural
modifications and transport properties in dense fluids.
\end{abstract}
\date{\today}
 \maketitle

With the recent advent of the Inelastic X-Ray technique (IXS)
\cite{BURKEL,sette_oldwater}, that complemented the Inelastic
Neutron Scattering one extending the accessible exchanged momentum
$Q$ to the whole first pseudo Brillouin zone, the collective
dynamics of disordered systems have been investigated in different
classes of liquids and glasses
\cite{scop_rmp,set_sci,ruo_nat,ruo_glassreview,sco_sci}. Among
other features, the specific interest here is the discovery that
in all the investigated liquids and glasses it exists a positive
dispersion of the sound speed, i.e. an increase of the velocity of
the longitudinal acoustic mode from the (relaxed) ultrasonic value
to an infinite frequency (unrelaxed) value. Such
\textit{anomalous} dispersion is ascribed to the presence of one
(or more) relaxation processes interacting with the dynamics of
the density fluctuations.

On a qualitative ground, a relaxation process is characterized by
a specific time, $\tau$, which marks the border between "viscous
liquid"-like dynamics (for $\omega \tau<<1$)  to "elastic
solid"-like one (for $\omega \tau>>1$). Correspondingly, the
longitudinal sound velocity, i.~e. the velocity of propagation of
the density fluctuation, undergoes a transition (positive
dispersion) from its "low" frequency limit, $c_0$, which
characterizes the "liquid" value, to its "infinite" frequency
limit $c_\infty > c_0$, characteristic of the "solid" response of
the system.

Usually, more than one single relaxation process is active, and a
"sequence" of sound velocity dispersions takes place. In
particular, convincing evidence has been reported in the last
years supporting a scenario in which the density fluctuations
decay according to three different relaxation mechanisms
\cite{scop_prlli}: {\it i)} a thermal process ($\tau_{th}$),
arising from the coupling of temperature and density, which marks
the transition from adiabatic ($\omega \tau_{th}>>1$) to
isothermal ($\omega \tau_{th}<<1$) dynamics. {\it ii)} a
structural (or $\alpha$) process, strongly temperature dependent,
whose timescale $\tau_\alpha$ is directly related to the viscosity
and hence to the mass diffusion coefficient. {\it iii)} a
microscopic process, which drives the decay of the density
fluctuations at a given wavelength, induced by the underlying
disordered structure that does not support this wavelength as an
energy eigenstate \cite{gcr_prlsim}. This scheme, validated by
several experimental and numerical evidences, has been developed
for glasses and normal liquids, i.e. for systems well below their
liquid-gas critical point.

In this letter we investigate whether this approach can be applied
to the supercritical phase, i.e. if and how the dynamics of a
dense liquid evolves on crossing the critical temperature. On a
qualitative ground, one might expect that on abandoning the liquid
phase the positive dispersion should vanish. As matter of fact
both the structural and microscopic processes are likely to become
inactive as soon as the dynamics loses its cooperative nature and
the concept of instantaneous vibrations is no longer well defined.
Looking at the available IXS literature data for simple liquids
and supercritical fluids
\cite{hos_hg,scop_prlli,scop_prena,mon_k,cop_rb,bod_cs,cun_ar,Cunsolo_Neon_PRL,Bencivenga_N2_EPL},
one may actually conclude that the positive dispersion vanishes on
approaching the critical temperature. However, this simplified
view seems to be contradicted by the results reported here for
supercritical Oxygen: at twice the critical temperature we still
find a 20 $\%$ positive dispersion. This result seems to indicate
that the end of the liquid dynamics is not only marked by the
critical temperature. Moreover, by examining the whole set of
existing data, we conclude that the extension of the liquid-vapor
coexistence line into the supercritical region (corresponding to
the Widom's line, i.e. the thermodynamic locus of the specific
heat maxima) defines the transition from liquid-like to gas-like
dynamics. This finding could help to give insights to the
phenomena of the non metal-metal transition in hot dense simple
fluids such as Hydrogen, Oxygen and Nitrogen, that has been
observed at thermodynamic points which actually lie on the
extrapolation of the coexistence line.

The experiment was carried out at the IXS Beamline II (ID28) at
the ESRF. The X-rays from an undulator are monochromatized by a
double crystal monochromator and a high-energy resolution
backscattering monochromator, operating at the silicon (9 9 9)
reflection order. The backscattered photons of energy 17794 eV
(3$\times$10$^{10}$ photon/s) are focused by a cylindrical mirror
and a Kirkpatrick-Baez multiplayer, providing a focal spot of 30
(horizontal) by 80 (vertical) $\mu$m$^{2}$ Full Width at Half
Maximum (FWHM) at the sample position. The scattered photons are
energy-analyzed by a Rowland circle five-crystal spectrometer,
with an overall energy resolution $\delta E \approx$ 3.0 meV, and
detected by a Peltier-cooled silicon diode.
The momentum transfer $Q = 2k_{i} \cdot sin(\theta_{s}/2) $, where
$k_i$ is the incident photon wave vector and $\theta_s$ is the
scattering angle, is selected by rotating the spectrometer around
a vertical axis passing through the scattering sample. Spectra at
five different momentum transfers can be recorded simultaneously
by five independent analyzer systems with a resolution $\delta Q
\approx 0.2$ nm$^{-1}$. We have used a diamond anvil cell (DAC) of
the membrane type, equipped with 600 $\mu$m culet diamonds. The
cell was loaded by condensing high purity liquid oxygen in a
sealed vessel at low temperature. Pressure is measured by the
shift of the ruby fluorescence wavelength \footnote{The
calibration curve, taken from Ref.~\cite{Calibrazione_rubino}, is
given by $P[GPa] = \lambda_0\times 0.3574\times [(\lambda
[nm]/\lambda _0)^{7.665}-1]$ where $\lambda_0$=694.28 nm.}. The
sample dimensions were about 300 $\mu$m in diameter, and 80 $\mu$m
in thickness. The IXS measurements were performed at room
temperature at 0.88, 2.88 and 5.35 GPa ($\pm 0.01$ GPa). The
highest pressure experimental point is close to the melting line,
but still in the fluid phase as was obtained by melting the
crystal along an isothermal decompression (300 K) under visual
inspection by using a microscope.

\begin{figure}[h]
\centering
\includegraphics[width=.48\textwidth]{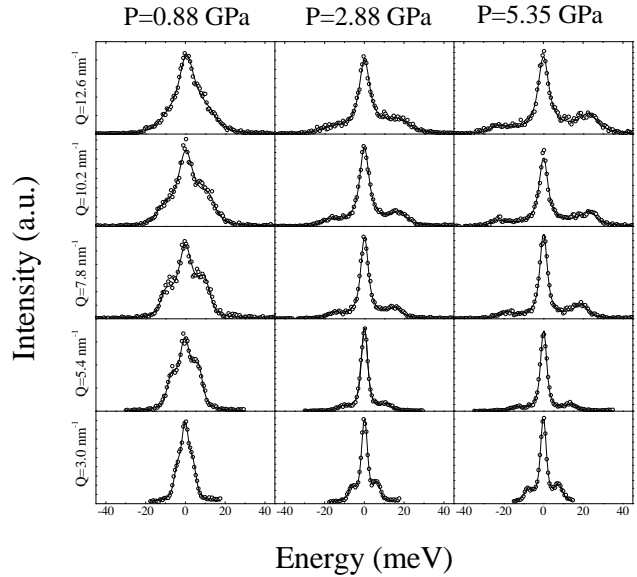}
\vspace{-3.7 cm}\caption{Selected IXS spectra of supercritical
oxygen at $T$=300 K. The three columns correspond to three
different pressure values (0.88, 2.88 and 5.35 GPa from left to
right) and the different rows report spectra taken at the
indicated $Q$ values (3.0, 5.4, 7.8, 10.2 and 12.6 nm$^{-1}$ from
bottom to top).} \label{f1}
\end{figure}

The evolution of the IXS spectra as a function of the momentum
transfer, $Q$, is shown in Fig.\ref{f1} for the $3.0 \div 12.6$
nm$^{-1}$ $Q$-range and for the three investigated pressure
values. The inelastic signal can be clearly observed at the two
sides of the quasi-elastic line: it is structured at the low
$Q$-values, and it becomes broad and featureless on increasing
$Q$, although it is still well visible on top of the tails of the
quasi-elastic peak.

The IXS spectra are related to the classical dynamic structure
factor $S(Q,\omega)$ through the detailed balance factor and the
instrumental resolution function, $R(\omega)$, through:

\begin{equation}
I(Q,\omega )=\int \frac{\hbar \omega ^{\prime }/KT}{1-e^{-\hbar
\omega ^{\prime }/k_BT}}S(Q,\omega ^{\prime })R(\omega -\omega
^{\prime })d\omega ^{\prime } \label{convo}
\end{equation}

A model for the classical $S(Q, \omega)$ has been obtained
exploiting the generalized hydrodynamic approach, in which the
dynamic structure factor is related to the second order memory
function through \cite{BALUCANI}:

\begin{equation}
\frac{S(Q,\omega )}{S(Q)}=\frac{\omega _0^2(Q)\tilde M^{\prime }(Q,\omega )/\pi}{%
\left[ \omega ^2-\omega _0^2-\omega \tilde M^{\prime \prime
}(Q,\omega )\right] ^2+\left[ \omega \tilde M^{\prime }(Q,\omega
)\right] ^2} \label{sqwgenerale}
\end{equation}

In this expression $\omega_0=k_BTQ^2/mS(Q)$, being $m$ the
molecular mass, $k_B$ the Boltzmann constant and $S(Q)$ the static
structure factor, and $\tilde M(Q,\omega)= \tilde M'(Q,\omega)+i
\tilde M''(Q,\omega)$ is the memory function in the frequency
domain. In the hydrodynamic limit, the decay of the memory
function proceeds via a timescale related to the  thermal
diffusivity $\tau_{th}=\gamma D_T Q^2$, where $\gamma$ is the
specific heat ratio. A second decay, whose timescale
($\tau_\alpha$) is related to the typical structural relaxation
time, is introduced to account for the viscoelastic nature of the
system. Finally, a third, microscopic, relaxation time is
introduced to represent the effect of the disordered instantaneous
atomic arrangement on the propagation of the density fluctuation
at a given Q. As in most cases \cite{scop_rmp}, this third
contribution can be considered as "instantaneous", and represented
by a delta-function. Consequently, the memory function in the time
domain reads:

\begin{equation}
M(Q,t)=(\gamma-1)\omega_o^2 e^{-\gamma D_TQ^2t} + 2 \Gamma Q^2
\delta(t) + \Delta _L^2 e^{-t/\tau_\alpha} \label{memoryth}
\end{equation}

In this expression $\Gamma Q^2$ is the amplitude of the
microscopic process, while $\Delta_L^2=(c_\infty^2-c_o^2)Q^2$ is
the intensity of the structural process, being $c_o$ and
$c_\infty$ the low and high frequency sound velocities
respectively . By plugging this expression into
Eq.~(\ref{sqwgenerale}), and using Eq.~(\ref{convo}), we were able
to describe the experimental data, as shown by the best fit
reported in Fig.~\ref{f1}. In the fitting procedure, the
parameters $\Delta$, $\tau_\alpha$, $\Gamma$ and $\omega_0$ were
left free, $D_T$ and $\gamma$ were fixed to their $Q$=0 values
obtained by the Impulsive Stimulated Thermal Scattering (ISTS)
data \cite{abram_SS,abram_TD}.

\begin{figure}[h]
\centering
\includegraphics[width=.46\textwidth]{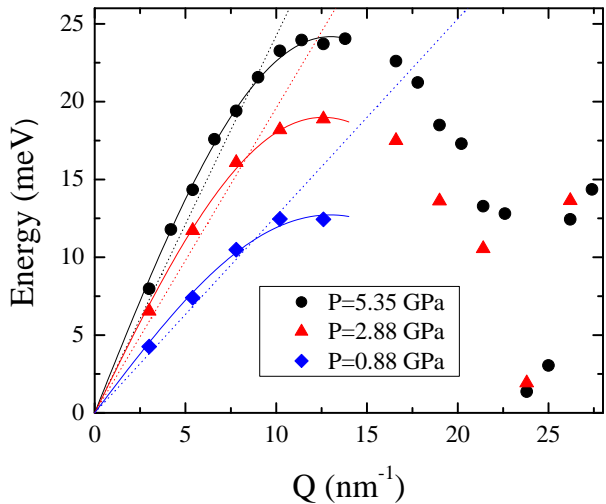}
\vspace{-4.4 cm}\caption{Dispersion relation of supercritical
oxygen at $T$=300 K at the three measured pressures (circles:
P=5.35 GPa, triangles: P=2.88 GPa, diamonds: P=0.88 GPa), from the
current maxima associated to the best fit according to
Eq.~(\ref{sqwgenerale}). The dotted lines indicate the adiabatic
sound velocities as determined by low frequency (GHz) ISTS
measurements \cite{abram_SS}, while the continuous lines are a
sine-like best fit whose $Q \rightarrow 0$ slope represents the
high frequency sound velocity.}
 \label{f2}
\end{figure}

\begin{table}[h]
\renewcommand{\arraystretch}{0.5}
\begin{tabular}{ccc} $P$ [GPa])&$c_{ISTS} [m/s]$&$c_{IXS} [m/s]$\\ \hline
$0.88$&1920&2340\\
2.88&2980&3600\\
5.35&3680&4440\\
 \hline
\end{tabular}
\vspace{.5 cm}\caption{Low and high frequency sound velocity in
supercritical oxygen. Experimental uncertainties are $\pm 50$
m/s.} \label{table}
\end{table}

In Fig.~\ref{f2} we report the dispersion curves as determined by
IXS at the three investigated pressures. These were obtained from
the maxima of the current correlation spectra $J(Q,\omega)=\omega
^2/Q^2 S(Q,\omega)$ (circles), utilizing the fit results for the
model $S(Q,\omega)$. For space reason the values of the other
fitting parameters are not reported here. For each pressure, the
low $Q$ slope of the evolution of the energy values represents the
sound velocity measured at high frequency. We note that the
observed high frequency sound velocity values always lie above the
low frequency, adiabatic sound velocity measured by the ISTS
technique \cite{abram_SS}(see Table 1) \footnote{Since above the
melting temperature both the structural and microscopic relaxation
times are below the picosecond, and the frequency of an ISTS
experiment is in the GHz range, the condition $\omega \tau << 1$
is always fulfilled. Hence, the adiabatic, low frequency sound
velocity is probed in this kind of experiments.}. This result
demonstrates the presence of positive dispersion -the fingerprint
of a liquid-like dynamics- in a supercritical fluid well above its
$T_c$ and $P_c$. As an immediate consequence, one has to rule out
the idea that the critical temperature alone marks the boundary of
simple liquid dynamics.

\begin{figure}[h]
\includegraphics[width=.5\textwidth]{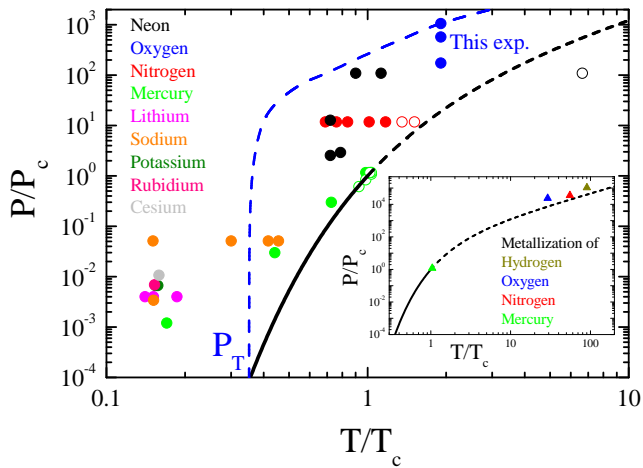}
\vspace{-6.5 cm}\caption{Sketch of the ($P/P_c$-$T/T_c$) plane.
The black lines are the best fit of the average of the
 liquid-vapour coexistence lines for Neon, Nitrogen and Oxygen
using Eq. \ref{coex} (see text), drawn below (continuous line) and
extrapolated above (dashed line) the critical point, respectively.
The dashed blue line is the melting line of Oxygen
\cite{GAS_ENCYCL,Young}. Each color corresponds to a different
investigated system. Open points represent cases where the
positive dispersion of the sound velocity has not been observed,
full points cases where there is a clear signature of positive
dispersion. Data for liquid metals are taken from \cite{scop_rmp}
and references therein (see Table I); data on Neon are from
\cite{Well_Neon_PRA} and from \cite{cun_ar,Cunsolo_Neon_PRL}; data
on
 Mercury are from \cite{inui_sn}; data on Nitrogen
are from \cite{Bencivenga_N2_EPL}. Inset: thermodynamic points
 where the non metal-metal transition has been experimentally observed
for Mercury (ref. \cite{Hensel_Hg_M-NM_REV_MOD_PHYS}), Oxygen
(ref. \cite{bastea}), Nitrogen (ref. \cite{chau}), and Hydrogen
(ref. \cite{weir}). The last three cases were investigated by
means of shock wave experiments.}
 \label{f3}
\end{figure}

In order to asses what really determines the end of the simple
liquid dynamics regime, we compared the present results for oxygen
to existing literature data. In fig~\ref{f3} we highlight the
presence of positive dispersion in a library of fluids at several
thermodynamic points. Full and open circles indicate a positive
dispersion above and below 5$\%$, respectively. As it can be seen,
systems around or moderately below the critical temperature, but
well above $P_c$ still exhibit positive dispersion, which fades
out on increasing temperature. What is really surprising is that
positive dispersion still persists in a strongly supercritical
system, the case of Oxygen reported here, in which both
temperature and pressure are well above the critical point. In
order to rationalize such a puzzling behaviour, we report in
Fig.~\ref{f3} the Plank-Riedel equation for the coexistence line
\cite{plankriedel}:

\begin{equation}
ln \left (\frac{P}{P_c}\right )=a+b\frac{T_c}{T}+cln \left
(\frac{T}{T_c}\right ) \label{coex}
\end{equation}

The values of a, b and c (a=4.270, b=-4.271 and c=1.414) have been
obtained by fitting the liquid-vapour coexistence lines of Neon,
Oxygen and Nitrogen \cite{GAS_ENCYCL}.  Eq.~\ref{coex} can
represent the noble gases liquid-vapour coexistence line
\cite{GAS_ENCYCL} in the spirit of the principle of corresponding
states, but it also describes rather well the liquid-vapour
coexistence line for the other simple fluids discussed here (see
Fig.~\ref{f3}). This curve is extrapolated here above the critical
point. On the basis of Fig.~\ref{f3}, one is led to conclude that
the end of simple liquid dynamics is identified by the
extrapolation of the coexistence line although, on a strict
thermodynamical ground, such line has a well defined meaning only
below the critical point.

Recent shock wave measurements have shown a density and
temperature driven non metal-metal transition in hot dense fluid
Oxygen \cite{bastea}, Nitrogen \cite{chau} and Hydrogen
\cite{weir} at thermodynamic points which surprisingly lie on the
proposed extrapolation line, much above the region where acoustic
properties have been investigated (inset of figure \ref{f3}). Also
in the case of Mercury, which is a metallic liquid and an
insulator gas \cite{Hensel_Hg_M-NM_REV_MOD_PHYS}, the non metal-
metal transition in the supercritical fluid, observed slightly
above the critical point, takes place around the extrapolation of
the liquid-vapour coexistence line, see Fig.~\ref{f3}. As a matter
of fact, the metal-non metal transition of expanded mercury, which
has been extensively studied \cite{inui_sn} and references
therein), appears to be related to the density decrease which
takes place on crossing the extrapolation of the liquid-vapour
coexistence line. Furthermore the positive dispersion disappears
on crossing the same extrapolation, as shown in Fig.~\ref{f3}, so
that in this case we have the exact coincidence of the two
phenomena: the change in dynamics and the metallization
transition.

In conclusion, we have shown here that a strongly supercritical
fluid (Oxygen at $T/T_c \approx 2$ and $P/P_c > 100$) exhibits
dynamical properties that are commonly ascribed to ordinary liquids.
Based on the comparison with the behavior of several other fluids
as a function of their thermodynamic state, we conclude that the
extrapolation well above the critical point of the liquid-vapor
coexistence line marks the boundary between simple liquid
dynamics and the collision dominated regime characterizing the gas
phase. Furthermore, the existing data on the location of the non metal-metal
transition of mercury and of hot dense simple fluids obtained by shock wave
 experiments indicate that the structural changes which take place on crossing
the extrapolation of the liquid-vapor coexistence line are also
responsible for
 the non metal-metal transition in overcritical fluids.

We suggest that the extension of the coexistence line beyond the
critical point splits the P-T phase diagram of simple fluids into
a gas-like and a liquid-like domain, respectively, which differ in
the
 local structure, resembling the sub-critical behaviors. Specifically,
 the "liquid-like phase" is expected to be denser and less rich in entropy than the
"gas-like phase". These differences, even if subtle at a first
glance, could support the major changes observed in the dynamical
and transport properties. If this point will be confirmed by
further experiments on different fluids and on an extended
thermodynamic region, it would imply a profound interconnection
between structural modifications in fluid materials occurring at
very different P-T conditions which, in turn, could stimulate and
support a new extended and unified view of fluid matter
thermodynamics.

We acknowledge Alexandre Beraud for assisting us during the
experiment.


\begin{thebibliography}{30}
\expandafter\ifx\csname
natexlab\endcsname\relax\def\natexlab#1{#1}\fi
\expandafter\ifx\csname bibnamefont\endcsname\relax
  \def\bibnamefont#1{#1}\fi
\expandafter\ifx\csname bibfnamefont\endcsname\relax
  \def\bibfnamefont#1{#1}\fi
\expandafter\ifx\csname citenamefont\endcsname\relax
  \def\citenamefont#1{#1}\fi
\expandafter\ifx\csname url\endcsname\relax
  \def\url#1{\texttt{#1}}\fi
\expandafter\ifx\csname
urlprefix\endcsname\relax\def\urlprefix{URL }\fi
\providecommand{\bibinfo}[2]{#2}
\providecommand{\eprint}[2][]{\url{#2}}

\bibitem[{\citenamefont{Burkel}(1991)}]{BURKEL}
\bibinfo{author}{\bibfnamefont{E.}~\bibnamefont{Burkel}},
  \emph{\bibinfo{title}{Inelastic Scattering of X-rays with very high Energy
  Resolution}} (\bibinfo{publisher}{Springer Verlag, Berlin},
  \bibinfo{year}{1991}).

\bibitem[{\citenamefont{Sette et~al.}(1995)\citenamefont{Sette, Ruocco, Krisch,
  Bergmann, Masciovecchio, Mazzacurati, Signorelli, and
  Verbeni}}]{sette_oldwater}
\bibinfo{author}{\bibfnamefont{F.}~\bibnamefont{Sette}},
  \bibinfo{author}{\bibfnamefont{G.}~\bibnamefont{Ruocco}},
  \bibinfo{author}{\bibfnamefont{M.}~\bibnamefont{Krisch}},
  \bibinfo{author}{\bibfnamefont{U.}~\bibnamefont{Bergmann}},
  \bibinfo{author}{\bibfnamefont{C.}~\bibnamefont{Masciovecchio}},
  \bibinfo{author}{\bibfnamefont{V.}~\bibnamefont{Mazzacurati}},
  \bibinfo{author}{\bibfnamefont{G.}~\bibnamefont{Signorelli}},
  \bibnamefont{and} \bibinfo{author}{\bibfnamefont{R.}~\bibnamefont{Verbeni}},
  \bibinfo{journal}{Phys. Rev. Lett.} \textbf{\bibinfo{volume}{75}},
  \bibinfo{pages}{850} (\bibinfo{year}{1995}).

\bibitem[{\citenamefont{Scopigno et~al.}(2005)\citenamefont{Scopigno, Ruocco,
  and Sette}}]{scop_rmp}
\bibinfo{author}{\bibfnamefont{T.}~\bibnamefont{Scopigno}},
  \bibinfo{author}{\bibfnamefont{G.}~\bibnamefont{Ruocco}}, \bibnamefont{and}
  \bibinfo{author}{\bibfnamefont{F.}~\bibnamefont{Sette}},
  \bibinfo{journal}{Rev. Mod. Phys.} \textbf{\bibinfo{volume}{77}},
  \bibinfo{pages}{881} (\bibinfo{year}{2005}).

\bibitem[{\citenamefont{Sette et~al.}(1998)\citenamefont{Sette, Krisch,
  Masciovecchio, Ruocco, and Monaco}}]{set_sci}
\bibinfo{author}{\bibfnamefont{F.}~\bibnamefont{Sette}},
  \bibinfo{author}{\bibfnamefont{M.}~\bibnamefont{Krisch}},
  \bibinfo{author}{\bibfnamefont{C.}~\bibnamefont{Masciovecchio}},
  \bibinfo{author}{\bibfnamefont{G.}~\bibnamefont{Ruocco}}, \bibnamefont{and}
  \bibinfo{author}{\bibfnamefont{G.}~\bibnamefont{Monaco}},
  \bibinfo{journal}{Science} \textbf{\bibinfo{volume}{280}},
  \bibinfo{pages}{1550} (\bibinfo{year}{1998}).

\bibitem[{\citenamefont{Ruocco et~al.}(1996)\citenamefont{Ruocco, Sette,
  Krisch, Masciovecchio, Mazzacurati, Signorelli, and Verbeni.}}]{ruo_nat}
\bibinfo{author}{\bibfnamefont{G.}~\bibnamefont{Ruocco}},
  \bibinfo{author}{\bibfnamefont{F.}~\bibnamefont{Sette}},
  \bibinfo{author}{\bibfnamefont{M.}~\bibnamefont{Krisch}},
  \bibinfo{author}{\bibfnamefont{U.~B.~C.} \bibnamefont{Masciovecchio}},
  \bibinfo{author}{\bibfnamefont{V.}~\bibnamefont{Mazzacurati}},
  \bibinfo{author}{\bibfnamefont{G.}~\bibnamefont{Signorelli}},
  \bibnamefont{and} \bibinfo{author}{\bibfnamefont{R.}~\bibnamefont{Verbeni.}},
  \bibinfo{journal}{Nature} \textbf{\bibinfo{volume}{379}},
  \bibinfo{pages}{521} (\bibinfo{year}{1996}).

\bibitem[{\citenamefont{Ruocco and Sette}(2001)}]{ruo_glassreview}
\bibinfo{author}{\bibfnamefont{G.}~\bibnamefont{Ruocco}} \bibnamefont{and}
  \bibinfo{author}{\bibfnamefont{F.}~\bibnamefont{Sette}}, \bibinfo{journal}{J.
  Phys. C} \textbf{\bibinfo{volume}{13}}, \bibinfo{pages}{9141}
  (\bibinfo{year}{2001}).

\bibitem[{\citenamefont{Scopigno et~al.}(2003)\citenamefont{Scopigno, Ruocco,
  Sette, and Monaco}}]{sco_sci}
\bibinfo{author}{\bibfnamefont{T.}~\bibnamefont{Scopigno}},
  \bibinfo{author}{\bibfnamefont{G.}~\bibnamefont{Ruocco}},
  \bibinfo{author}{\bibfnamefont{F.}~\bibnamefont{Sette}}, \bibnamefont{and}
  \bibinfo{author}{\bibfnamefont{G.}~\bibnamefont{Monaco}},
  \bibinfo{journal}{Science} \textbf{\bibinfo{volume}{302}},
  \bibinfo{pages}{849} (\bibinfo{year}{2003}).

\bibitem[{\citenamefont{Scopigno et~al.}(2000)\citenamefont{Scopigno, Balucani,
  Ruocco, and Sette}}]{scop_prlli}
\bibinfo{author}{\bibfnamefont{T.}~\bibnamefont{Scopigno}},
  \bibinfo{author}{\bibfnamefont{U.}~\bibnamefont{Balucani}},
  \bibinfo{author}{\bibfnamefont{G.}~\bibnamefont{Ruocco}}, \bibnamefont{and}
  \bibinfo{author}{\bibfnamefont{F.}~\bibnamefont{Sette}},
  \bibinfo{journal}{Phys. Rev. Lett.} \textbf{\bibinfo{volume}{85}},
  \bibinfo{pages}{4076} (\bibinfo{year}{2000}).

\bibitem[{\citenamefont{Ruocco et~al.}(2000)\citenamefont{Ruocco, Sette, {Di
  Leonardo}, Monaco, Sampoli, Scopigno, and Viliani}}]{gcr_prlsim}
\bibinfo{author}{\bibfnamefont{G.}~\bibnamefont{Ruocco}},
  \bibinfo{author}{\bibfnamefont{F.}~\bibnamefont{Sette}},
  \bibinfo{author}{\bibfnamefont{R.}~\bibnamefont{{Di Leonardo}}},
  \bibinfo{author}{\bibfnamefont{G.}~\bibnamefont{Monaco}},
  \bibinfo{author}{\bibfnamefont{M.}~\bibnamefont{Sampoli}},
  \bibinfo{author}{\bibfnamefont{T.}~\bibnamefont{Scopigno}}, \bibnamefont{and}
  \bibinfo{author}{\bibfnamefont{G.}~\bibnamefont{Viliani}},
  \bibinfo{journal}{Phys. Rev. Lett.} \textbf{\bibinfo{volume}{84}},
  \bibinfo{pages}{5788} (\bibinfo{year}{2000}).

\bibitem[{\citenamefont{Hosokawa et~al.}(2002)\citenamefont{Hosokawa, Sinn,
  Hensel, Alatas, Alp, and Pilgrim}}]{hos_hg}
\bibinfo{author}{\bibfnamefont{S.}~\bibnamefont{Hosokawa}},
  \bibinfo{author}{\bibfnamefont{H.}~\bibnamefont{Sinn}},
  \bibinfo{author}{\bibfnamefont{F.}~\bibnamefont{Hensel}},
  \bibinfo{author}{\bibfnamefont{A.}~\bibnamefont{Alatas}},
  \bibinfo{author}{\bibfnamefont{E.~E.} \bibnamefont{Alp}}, \bibnamefont{and}
  \bibinfo{author}{\bibfnamefont{W.-C.} \bibnamefont{Pilgrim}},
  \bibinfo{journal}{J. Non-Cryst. Solids} \textbf{\bibinfo{volume}{312-314}},
  \bibinfo{pages}{163} (\bibinfo{year}{2002}).

\bibitem[{\citenamefont{Scopigno et~al.}(2002)\citenamefont{Scopigno, Balucani,
  Ruocco, and Sette}}]{scop_prena}
\bibinfo{author}{\bibfnamefont{T.}~\bibnamefont{Scopigno}},
  \bibinfo{author}{\bibfnamefont{U.}~\bibnamefont{Balucani}},
  \bibinfo{author}{\bibfnamefont{G.}~\bibnamefont{Ruocco}}, \bibnamefont{and}
  \bibinfo{author}{\bibfnamefont{F.}~\bibnamefont{Sette}},
  \bibinfo{journal}{Phys. Rev. E} \textbf{\bibinfo{volume}{65}},
  \bibinfo{pages}{031205} (\bibinfo{year}{2002}).

\bibitem[{\citenamefont{Monaco et~al.}(2004)\citenamefont{Monaco, Scopigno,
  Benassi, Giugni, Monaco, Nardone, Ruocco, and Sampoli}}]{mon_k}
\bibinfo{author}{\bibfnamefont{A.}~\bibnamefont{Monaco}},
  \bibinfo{author}{\bibfnamefont{T.}~\bibnamefont{Scopigno}},
  \bibinfo{author}{\bibfnamefont{P.}~\bibnamefont{Benassi}},
  \bibinfo{author}{\bibfnamefont{A.}~\bibnamefont{Giugni}},
  \bibinfo{author}{\bibfnamefont{G.}~\bibnamefont{Monaco}},
  \bibinfo{author}{\bibfnamefont{M.}~\bibnamefont{Nardone}},
  \bibinfo{author}{\bibfnamefont{G.}~\bibnamefont{Ruocco}}, \bibnamefont{and}
  \bibinfo{author}{\bibfnamefont{M.}~\bibnamefont{Sampoli}},
  \bibinfo{journal}{J. Chem. Phys.} \textbf{\bibinfo{volume}{120}},
  \bibinfo{pages}{8089} (\bibinfo{year}{2004}).

\bibitem[{\citenamefont{Copley and Rowe}(1974)}]{cop_rb}
\bibinfo{author}{\bibfnamefont{J.~R.~D.} \bibnamefont{Copley}}
  \bibnamefont{and} \bibinfo{author}{\bibfnamefont{M.}~\bibnamefont{Rowe}},
  \bibinfo{journal}{Phys. Rev. A} \textbf{\bibinfo{volume}{9}},
  \bibinfo{pages}{1656} (\bibinfo{year}{1974}).

\bibitem[{\citenamefont{Bodensteiner et~al.}(1992)\citenamefont{Bodensteiner,
  Morkel, Gl{\"a}ser, and Dorner}}]{bod_cs}
\bibinfo{author}{\bibfnamefont{T.}~\bibnamefont{Bodensteiner}},
  \bibinfo{author}{\bibfnamefont{C.}~\bibnamefont{Morkel}},
  \bibinfo{author}{\bibfnamefont{W.}~\bibnamefont{Gl{\"a}ser}},
  \bibnamefont{and} \bibinfo{author}{\bibfnamefont{B.}~\bibnamefont{Dorner}},
  \bibinfo{journal}{Phys. Rev. A} \textbf{\bibinfo{volume}{45}},
  \bibinfo{pages}{5709} (\bibinfo{year}{1992}).

\bibitem[{\citenamefont{Cunsolo et~al.}(2001)\citenamefont{Cunsolo, Pratesi,
  Verbeni, Colognesi, Masciovecchio, Monaco, Ruocco, and Sette}}]{cun_ar}
\bibinfo{author}{\bibfnamefont{A.}~\bibnamefont{Cunsolo}},
  \bibinfo{author}{\bibfnamefont{G.}~\bibnamefont{Pratesi}},
  \bibinfo{author}{\bibfnamefont{R.}~\bibnamefont{Verbeni}},
  \bibinfo{author}{\bibfnamefont{D.}~\bibnamefont{Colognesi}},
  \bibinfo{author}{\bibfnamefont{C.}~\bibnamefont{Masciovecchio}},
  \bibinfo{author}{\bibfnamefont{G.}~\bibnamefont{Monaco}},
  \bibinfo{author}{\bibfnamefont{G.}~\bibnamefont{Ruocco}}, \bibnamefont{and}
  \bibinfo{author}{\bibfnamefont{F.}~\bibnamefont{Sette}}, \bibinfo{journal}{J.
  Chem. Phys.} \textbf{\bibinfo{volume}{114}}, \bibinfo{pages}{2259}
  (\bibinfo{year}{2001}).

\bibitem[{\citenamefont{{Cunsolo} et~al.}(1998)\citenamefont{{Cunsolo},
  Pratesi, Ruocco, Sampoli, Sette, Verbeni, Barocchi, Krisch, Masciovecchio,
  and Nardone}}]{Cunsolo_Neon_PRL}
\bibinfo{author}{\bibfnamefont{A.}~\bibnamefont{{Cunsolo}}},
  \bibinfo{author}{\bibfnamefont{G.}~\bibnamefont{Pratesi}},
  \bibinfo{author}{\bibfnamefont{G.}~\bibnamefont{Ruocco}},
  \bibinfo{author}{\bibfnamefont{M.}~\bibnamefont{Sampoli}},
  \bibinfo{author}{\bibfnamefont{F.}~\bibnamefont{Sette}},
  \bibinfo{author}{\bibfnamefont{R.}~\bibnamefont{Verbeni}},
  \bibinfo{author}{\bibfnamefont{F.}~\bibnamefont{Barocchi}},
  \bibinfo{author}{\bibfnamefont{M.}~\bibnamefont{Krisch}},
  \bibinfo{author}{\bibfnamefont{C.}~\bibnamefont{Masciovecchio}},
  \bibnamefont{and} \bibinfo{author}{\bibfnamefont{M.}~\bibnamefont{Nardone}},
  \bibinfo{journal}{Phys. Rev. Lett.} \textbf{\bibinfo{volume}{80}},
  \bibinfo{pages}{3515} (\bibinfo{year}{1998}).

\bibitem[{\citenamefont{{Bencivenga} et~al.}(2006)\citenamefont{{Bencivenga},
  Cunsolo, Krisch, Monaco, Ruocco, and Sette}}]{Bencivenga_N2_EPL}
\bibinfo{author}{\bibfnamefont{F.}~\bibnamefont{{Bencivenga}}},
  \bibinfo{author}{\bibfnamefont{A.}~\bibnamefont{Cunsolo}},
  \bibinfo{author}{\bibfnamefont{M.}~\bibnamefont{Krisch}},
  \bibinfo{author}{\bibfnamefont{G.}~\bibnamefont{Monaco}},
  \bibinfo{author}{\bibfnamefont{G.}~\bibnamefont{Ruocco}}, \bibnamefont{and}
  \bibinfo{author}{\bibfnamefont{F.}~\bibnamefont{Sette}},
  \bibinfo{journal}{Europhys. Lett.} \textbf{\bibinfo{volume}{00}},
  \bibinfo{pages}{0000} (\bibinfo{year}{2006}).

\bibitem[{\citenamefont{Balucani and Zoppi}(1983)}]{BALUCANI}
\bibinfo{author}{\bibfnamefont{U.}~\bibnamefont{Balucani}} \bibnamefont{and}
  \bibinfo{author}{\bibfnamefont{M.}~\bibnamefont{Zoppi}},
  \emph{\bibinfo{title}{Dynamics of the liquid state}}
  (\bibinfo{publisher}{Clarendon Press, Oxford}, \bibinfo{year}{1983}).

\bibitem[{\citenamefont{Abramson
  et~al.}(1999{\natexlab{a}})\citenamefont{Abramson, Slutsky, Harrell, and
  Brown}}]{abram_SS}
\bibinfo{author}{\bibfnamefont{E.}~\bibnamefont{Abramson}},
  \bibinfo{author}{\bibfnamefont{L.}~\bibnamefont{Slutsky}},
  \bibinfo{author}{\bibfnamefont{M.}~\bibnamefont{Harrell}}, \bibnamefont{and}
  \bibinfo{author}{\bibfnamefont{J.}~\bibnamefont{Brown}}, \bibinfo{journal}{J.
  Chem. Phys.} \textbf{\bibinfo{volume}{110}}, \bibinfo{pages}{10493}
  (\bibinfo{year}{1999}{\natexlab{a}}).

\bibitem[{\citenamefont{Abramson
  et~al.}(1999{\natexlab{b}})\citenamefont{Abramson, Slutsky, and
  Brown}}]{abram_TD}
\bibinfo{author}{\bibfnamefont{E.}~\bibnamefont{Abramson}},
  \bibinfo{author}{\bibfnamefont{L.}~\bibnamefont{Slutsky}}, \bibnamefont{and}
  \bibinfo{author}{\bibfnamefont{J.}~\bibnamefont{Brown}}, \bibinfo{journal}{J.
  Chem. Phys.} \textbf{\bibinfo{volume}{111}}, \bibinfo{pages}{1999}
  (\bibinfo{year}{1999}{\natexlab{b}}).

\bibitem[{\citenamefont{Plank and Riedel}(1948)}]{plankriedel}
\bibinfo{author}{\bibfnamefont{R.}~\bibnamefont{Plank}} \bibnamefont{and}
  \bibinfo{author}{\bibfnamefont{L.}~\bibnamefont{Riedel}},
  \bibinfo{journal}{Ingenieur-Archiv.} \textbf{\bibinfo{volume}{XVI}},
  \bibinfo{pages}{255} (\bibinfo{year}{1948}).

\bibitem[{\citenamefont{Medard}(1976)}]{GAS_ENCYCL}
\bibinfo{author}{\bibfnamefont{L.}~\bibnamefont{Medard}},
  \emph{\bibinfo{title}{Gas Encyclopaedia}} (\bibinfo{publisher}{Elsevier},
  \bibinfo{year}{1976}).

\bibitem[{\citenamefont{Bastea et~al.}(2001)\citenamefont{Bastea, Mitchell, and
  Nellis}}]{bastea}
\bibinfo{author}{\bibfnamefont{M.}~\bibnamefont{Bastea}},
  \bibinfo{author}{\bibfnamefont{A.~C.} \bibnamefont{Mitchell}},
  \bibnamefont{and} \bibinfo{author}{\bibfnamefont{W.~J.}
  \bibnamefont{Nellis}}, \bibinfo{journal}{Phys. Rev. Lett.}
  \textbf{\bibinfo{volume}{86}}, \bibinfo{pages}{3108} (\bibinfo{year}{2001}).

\bibitem[{\citenamefont{Chau et~al.}(2003)\citenamefont{Chau, Mitchell, Minich,
  and Nellis}}]{chau}
\bibinfo{author}{\bibfnamefont{R.}~\bibnamefont{Chau}},
  \bibinfo{author}{\bibfnamefont{A.~C.} \bibnamefont{Mitchell}},
  \bibinfo{author}{\bibfnamefont{R.~W.} \bibnamefont{Minich}},
  \bibnamefont{and} \bibinfo{author}{\bibfnamefont{W.~J.}
  \bibnamefont{Nellis}}, \bibinfo{journal}{Phys. Rev. Lett.}
  \textbf{\bibinfo{volume}{90}}, \bibinfo{pages}{245501}
  (\bibinfo{year}{2003}).

\bibitem[{\citenamefont{Weir et~al.}(1996)\citenamefont{Weir, Mitchell, and
  Nellis}}]{weir}
\bibinfo{author}{\bibfnamefont{S.~T.} \bibnamefont{Weir}},
  \bibinfo{author}{\bibfnamefont{A.~C.} \bibnamefont{Mitchell}},
  \bibnamefont{and} \bibinfo{author}{\bibfnamefont{W.~J.}
  \bibnamefont{Nellis}}, \bibinfo{journal}{Phys. Rev. Lett.}
  \textbf{\bibinfo{volume}{76}}, \bibinfo{pages}{1860} (\bibinfo{year}{1996}).

\bibitem[{\citenamefont{{Hensel} and
  Frank}(1968)}]{Hensel_Hg_M-NM_REV_MOD_PHYS}
\bibinfo{author}{\bibfnamefont{F.}~\bibnamefont{{Hensel}}} \bibnamefont{and}
  \bibinfo{author}{\bibfnamefont{E.~U.} \bibnamefont{Frank}},
  \bibinfo{journal}{Rev. Mod. Phys.} \textbf{\bibinfo{volume}{40}},
  \bibinfo{pages}{697} (\bibinfo{year}{1968}).

\bibitem[{\citenamefont{Ishikawa et~al.}(2004)\citenamefont{Ishikawa, Inui,
  Matsuda, Tamura, Tsutsui, and Baron}}]{inui_sn}
\bibinfo{author}{\bibfnamefont{D.}~\bibnamefont{Ishikawa}},
  \bibinfo{author}{\bibfnamefont{M.}~\bibnamefont{Inui}},
  \bibinfo{author}{\bibfnamefont{K.}~\bibnamefont{Matsuda}},
  \bibinfo{author}{\bibfnamefont{K.}~\bibnamefont{Tamura}},
  \bibinfo{author}{\bibfnamefont{S.}~\bibnamefont{Tsutsui}}, \bibnamefont{and}
  \bibinfo{author}{\bibfnamefont{A.~Q.~R.} \bibnamefont{Baron}},
  \bibinfo{journal}{Phys. Rev. Lett.} \textbf{\bibinfo{volume}{93}},
  \bibinfo{pages}{097801} (\bibinfo{year}{2004}).

\bibitem[{\citenamefont{Young et~al.}(1987)\citenamefont{Young, {C.-S. Zha},
  Boehler, Yen, Nicol, Zinn, Schiferl, Kinkead, Hanson, and Pinnick}}]{Young}
\bibinfo{author}{\bibfnamefont{D.~A.} \bibnamefont{Young}},
  \bibinfo{author}{\bibnamefont{{C.-S. Zha}}},
  \bibinfo{author}{\bibfnamefont{R.}~\bibnamefont{Boehler}},
  \bibinfo{author}{\bibfnamefont{J.}~\bibnamefont{Yen}},
  \bibinfo{author}{\bibfnamefont{M.}~\bibnamefont{Nicol}},
  \bibinfo{author}{\bibfnamefont{A.~S.} \bibnamefont{Zinn}},
  \bibinfo{author}{\bibfnamefont{D.}~\bibnamefont{Schiferl}},
  \bibinfo{author}{\bibfnamefont{S.}~\bibnamefont{Kinkead}},
  \bibinfo{author}{\bibfnamefont{R.~C.} \bibnamefont{Hanson}},
  \bibnamefont{and} \bibinfo{author}{\bibfnamefont{D.~A.}
  \bibnamefont{Pinnick}}, \bibinfo{journal}{Phys. Rev. B}
  \textbf{\bibinfo{volume}{35}}, \bibinfo{pages}{5353} (\bibinfo{year}{1987}).

\bibitem[{\citenamefont{{van Well} and {de Graaf}}(1985)}]{Well_Neon_PRA}
\bibinfo{author}{\bibfnamefont{A.~A.} \bibnamefont{{van Well}}}
  \bibnamefont{and} \bibinfo{author}{\bibfnamefont{L.~A.} \bibnamefont{{de
  Graaf}}}, \bibinfo{journal}{Phys. Rev. A} \textbf{\bibinfo{volume}{32}},
  \bibinfo{pages}{2396} (\bibinfo{year}{1985}).

\bibitem[{\citenamefont{{Mao} et~al.}(1986)\citenamefont{{Mao}, Xu, and
  Bell.}}]{Calibrazione_rubino}
\bibinfo{author}{\bibfnamefont{H.~K.} \bibnamefont{{Mao}}},
  \bibinfo{author}{\bibfnamefont{J.}~\bibnamefont{Xu}}, \bibnamefont{and}
  \bibinfo{author}{\bibfnamefont{P.~M.} \bibnamefont{Bell.}},
  \bibinfo{journal}{J. Geophys. Res.} \textbf{\bibinfo{volume}{91}},
  \bibinfo{pages}{4673} (\bibinfo{year}{1986}).

\end{thebibliography}

\end{document}